\documentclass[aps,prl,tightenlines,preprintnumbers,amsmath,amssymb,superscriptaddress,nofootinbib,twocolumn]{revtex4}

\usepackage{latexsym,cancel,amssymb,amsmath}
\usepackage{multirow}
\usepackage{graphicx}
\usepackage{color}
\usepackage{dcolumn}
\usepackage{bm}
\usepackage{hyperref}
\usepackage{slashed}
\usepackage{multirow}

\def\be{\begin{equation}}
\def\ee{\end{equation}}
\def\bea{\begin{eqnarray}}
\def\eea{\end{eqnarray}}

\def\bbuildrel#1_#2^#3{\mathrel{\mathop{\kern 0pt#1}\limits_{#2}^{#3}}}
\def\slash#1{\setbox0=\hbox{$#1$}#1\hskip-\wd0\dimen0=5pt\advance
       \dimen0 by-\ht0\advance\dimen0 by\dp0\lower0.5\dimen0\hbox
         to\wd0{\hss\sl/\/\hss}}

\newcommand{\gae}{\lower 2pt \hbox{$\, \buildrel {\scriptstyle >}\over {\scriptstyle
\sim}\,$}}
\newcommand{\lae}{\lower 2pt \hbox{$\, \buildrel {\scriptstyle <}\over {\scriptstyle
\sim}\,$}}

\newcommand{\beq}{\begin{eqnarray}}
\newcommand{\eeq}{\end{eqnarray}}

\newcommand{\ba}{\begin{array}}
\newcommand{\ea}{\end{array}}

\long\def\symbolfootnote[#1]#2{\begingroup%
\def\thefootnote{\fnsymbol{footnote}}\footnote[#1]{#2}\endgroup}

\def\lsim{\mathrel{\rlap{\lower4pt\hbox{\hskip1pt$\sim$}}
    \raise1pt\hbox{$<$}}}         
\def\gsim{\mathrel{\rlap{\lower4pt\hbox{\hskip1pt$\sim$}}
    \raise1pt\hbox{$>$}}}         

\def\lsim{\:\raisebox{-0.5ex}{$\stackrel{\textstyle<}{\sim}$}\:}
\def\gsim{\:\raisebox{-0.5ex}{$\stackrel{\textstyle>}{\sim}$}\:}

\def\beq{\begin{equation}}
\def\eeq{\end{equation}}
\def\bea{\begin{eqnarray}}
\def\eea{\end{eqnarray}}
\def\to{\rightarrow}

\def\pp{\vskip\baselineskip\noindent}

\begin{document}
\DeclareGraphicsExtensions{.jpg,.pdf,.mps,.png,}

\title{Unraveling The Physics Behind Modified Higgs Couplings -- \\   LHC vs. a Higgs Factory}

\author{Steven B. Giddings}
\affiliation{Department of Physics, University of California,
Santa Barbara, CA 93106, USA}

\author{Tao Liu}
\affiliation{Department of Physics, University of California,
Santa Barbara, CA 93106, USA}

\author{Ian Low}
\affiliation{High Energy Physics Division, Argonne National Laboratory, Argonne, IL 60439, USA}
\affiliation{Department of Physics and Astronomy, Northwestern University, Evanston, IL 60208, USA}
\affiliation{Kavli Institute for Theoretical Physics, University of California, Santa Barbara, CA 93106, USA}

\author{Eric Mintun}
\affiliation{Department of Physics, University of California,
Santa Barbara, CA 93106, USA}

\begin{abstract}
Strongly modified $h\gamma\gamma$ and $hgg$ couplings indicate new electroweak  and color mediators, respectively, with a light mass and a significant coupling to the Higgs boson. We point out  the Higgs boson could have a significant decay width into the mediators and propose uncovering the hidden new physics through such exotic decays, which can probe the Higgs coupling with the mediators directly. Focusing on the electroweak mediators, we study a simplified model using as an example final states with 
tau leptons and neutrinos.
Because one of the mediators is off-shell and its decay products are extremely soft, it is challenging to make a discovery at the Large Hadron Collider. A Higgs factory such as the International Linear Collider, however, could serve as a discovery machine for the EW mediators even in an early stage. 
\end{abstract}

\maketitle


\pp
{\bf Introduction} -- On July 4th, 2012 CERN announced the observation of a Higgs-like boson at the Large Hadron Collider (LHC) with a mass at around 126 GeV \cite{:2012gk,:2012gu}. Preliminary results based on  decay branching ratios indicate a genuine Higgs boson, not an imposter \cite{Low:2012rj}, while signal strengths in all observed channels are also consistent with those expected from a Standard Model (SM) Higgs boson, except in the diphoton channel where  early data  suggest an enhancement over the SM rate of ${\cal O}$(50\%)~\cite{atlashiggscombo, cmshiggscombo}.

The enhanced event rate in the diphoton channel could arise from modifying any of the following  quantities: 1) the production cross-section, which at leading order comes from the gluon fusion process, 2) the total width, which is dominated by the partial width of Higgs to $b\bar{b}$ \cite{Carena:2002qg,Carena:2011aa}, and 3) the partial width of Higgs to diphoton~\cite{Carena:2011aa,Carena:2012gp,Carena:2012xa,An:2012vp}. 1) and 2)  alter the signal rate in {\em all}  channels, while 3)  only affects the diphoton channel. Current experimental fits  favor a SM Higgs-gluon-gluon ($hgg$) coupling and an enhanced Higgs-to-diphoton ($h\gamma\gamma$)  coupling \cite{cmshiggscombo, atlashiggscoupling}, although the statistics is  limited and uncertainty quite large.

The $h\gamma\gamma$ and $hgg$  couplings  are of special importance. On the experimental side, these couplings enter into the $gg\to h\to \gamma\gamma$ channel, which is the main discovery channel of the Higgs boson at the LHC. On the theoretical side, both couplings are induced only at the loop-level and serve as indirect probes of any new particles with a significant coupling to the Higgs \cite{Low:2009di,Carena:2011aa,Carena:2012gp,Carena:2012xa,An:2012vp}.  In particular, new electroweak (colored) particles coupling to the Higgs would necessarily modify the $h\gamma\gamma$ ($hgg$) couplings. More importantly, if electroweak (EW) symmetry breaking is natural,  new particles with significant couplings to the Higgs must exist to soften the quadratic divergences in the Higgs mass. As a result, there are intricate connections between  modifications in the $h\gamma\gamma$ and $hgg$ couplings and the naturalness of TeV scale physics \cite{Carena:2012xa,Low:2009di}.

It was shown in Ref.~\cite{Carena:2011aa,Carena:2012gp,Carena:2012xa,An:2012vp} that a possible strong enhancement of the $h\gamma\gamma$ coupling indicates new EW states that  i) are light, on the order of a few hundreds GeV, and ii) couple to the Higgs boson significantly. Therefore, if the enhancement persists in the future,  a top priority will be to devise strategies to search for these light ``EW mediators'' and to probe their couplings with the Higgs boson. (There are ways to hide these new light states from direct search and precision EW constraints by, for example,  assigning a new "parity" to the new particles \cite{Carena:2012xa}.) A smoking gun signal of EW mediators is a modified rate in  Higgs decays into $Z\gamma$ final states \cite{Gainer:2011aa}, which correlates with deviations from the SM width in the diphoton channel \cite{Carena:2011aa,Carena:2012gp,Carena:2012xa,An:2012vp}. Another indirect probe lies in electroweak production of the mediators at the LHC, with search for their decays into SM particles \cite{Carena:2012gp,An:2012vp,ArkaniHamed:2012kq}. The former however can not identify the mediators directly while the latter does not involve couplings between the Higgs boson and the mediators.

In this letter we propose searching for the new physics behind the modified $h\gamma\gamma$ coupling via exotic Higgs decays. If the mediators decay to light particles, an on-shell Higgs can decay to the mediators, at least one of which is off-shell, much like Higgs decays to off-shell $W/Z$ bosons. The partial width of this exotic Higgs decay depends on the available modes and phase space for the subsequent mediator decays, which can only be computed in a specific model. However, it could be significant, especially in the parameter region giving rise to a strongly modified $h\gamma\gamma$ coupling where the mediators are light and couple to the Higgs significantly. A similar strategy can be applied for studying the $hgg$ coupling, if a strong modification is indicated by the LHC measurements in the near future.  

To illustrate this strategy, we will work in a simplified model with a EW scalar mediator, $\phi$, assuming for example that it mainly decays into tau and tau neutrino. One implementation of this is the MSSM with a gauged-$U(1)_{\rm PQ}$ extension~\cite{An:2012vp}, where the diphoton width can be enhanced either by EW vector-like fermions which are required for the $U(1)_{\rm PQ}$ anomaly cancellation or by their superpartners. These charged mediators can decay to SM particles and (or) their superpartiners and hence avoid overproduction in the early Universe. We will see that, given a Higgs mass $\sim126$ GeV, one of the mediators is very off-shell and it is difficult to search for such decays at the LHC, although with some optimistic assumptions it might be feasible. We then turn to a Higgs factory such as  the International Linear Collider (ILC) with $\sqrt{s} = 250$ GeV and show that the discovery potential is fairly promising. Therefore the Higgs factory can be not only a precision machine, but also a discovery machine for the truth behind the strongly modified $h\gamma\gamma$ or $hgg$ couplings.

\pp
{\bf Exotic  Decay Width} -- The partial decay width of $h\to \phi  \phi $ depends on three physical parameters: the mass of the scalar mediator, $m_{ \phi }$, its total width, $\Gamma_{ \phi }$, and its coupling with the Higgs, $c_{ \phi }$, which is defined as in $c_{ \phi } v\, h { \phi} { \phi}^\dagger$ with $v\approx 246$ GeV being the Higgs vacuum expectation value. As a comparison, the change in the $h\gamma\gamma$ coupling depends on two physical parameters: $m_{ \phi}$ and $c_{ \phi}$ \cite{Carena:2012xa}. 

Extending the calculations in~\cite{Low:2010jp}, it is easy to find the partial decay width of $h\to \phi \phi$ 
 \be
 \label{eq:offtotalwidth}
 \Gamma_{h\to \phi \phi} = \int_0^{u_1} dm_{+}^2 \int_0^{u_2} dm_{-}^2\,
  \frac{d\Gamma_{h\to \phi \phi}}{dm_{+}^2dm_{-}^2} \ .
 \ee
 where $m_\pm$ is the invariant mass of $ \phi_\pm$, $u_1=m_h^2$, $u_2=(m_h-m_{+})^2$, and
 \bea
&& \frac{d\Gamma_{h\to \phi \phi}}{dm_{+}^2dm_{-}^2}  =  \frac{c_{ \phi}^2 v^2}{16\pi m_h^3} \sqrt{ m_h^2-{(m_{+}+m_{-})^2}}\nonumber \\
&&\quad \times \sqrt{ m_h^2-{(m_{+}-m_{-})^2}} 
\, P_+ P_- \ .
\eea
$P_\pm$ are the propagators of ${\phi}^\pm$:
\be
P_\pm =  \frac{m_{\pm}\Gamma_{\pm}}{\pi}\frac{1}
{(m_\pm^2-m_{\phi}^2)^2+m_{\pm}^2\Gamma_{\pm}^2}\ .
\ee
Note here $m_{\pm}$ is the invariant mass of the possibly off-shell $\phi$, and $\Gamma_{\pm}$ is the total width of $\phi$ at the corresponding $m_{\pm}$, which is model-independent. For illustration, we consider a simplified model, assuming $\phi \to \tau + \nu_\tau$ predominantly. This was suggested in~\cite{An:2012vp}.  Then the partial width $\Gamma_\pm$ is 
\begin{eqnarray}
\Gamma_\pm = 
\frac{c_{\phi  \nu_\tau \tau}^2 m_\pm}{16
  \pi }  \ ,
\end{eqnarray}
where $c_{\phi  \nu_\tau \tau}$ is the $\phi  \nu_\tau \tau$ coupling in the mass eigenbasis. In this analysis, we assume $c_{\phi  \nu_\tau \tau} =0.6$, comparable with the EW coupling. For $m_{\phi}\sim 100$ GeV, the on-shell decay width is then 
\be
\Gamma_{\phi} \sim 0.7 \  \rm{GeV}  \ .
\ee

\begin{figure}[t]
\centering
\includegraphics[width=0.35\textwidth]{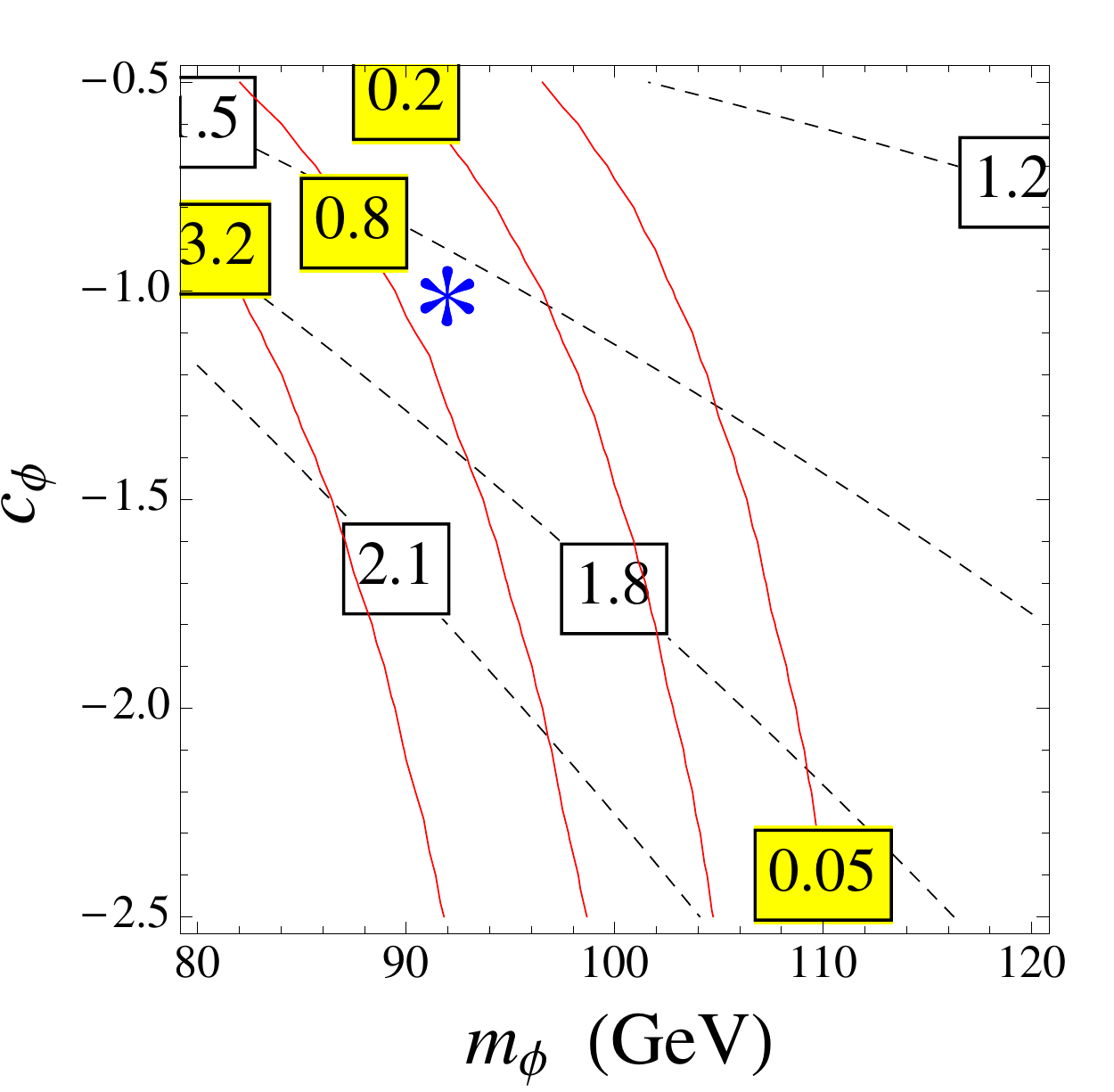}
\caption{Contours for ${\Gamma_{h\to \phi\phi}}/{\Gamma^{\rm SM}_{h\to \tau\bar \tau}}$ (red, solid) and ${\Gamma_{h\to \gamma\gamma}}/{\Gamma^{\rm SM}_{h\to \gamma\gamma}}$ (black, dashed). }
\label{mec1}
\end{figure}

The contours of $ \Gamma_{h\to \phi \phi}/\Gamma_{h\to \tau \bar \tau}^{\rm SM}$ and $\Gamma_{h\to \gamma\gamma}/\Gamma^{\rm SM}_{h\to \gamma\gamma}$ are shown in Fig.~\ref{mec1}. These contours indicate a strong positive correlation between $ \Gamma_{h\to \phi \phi}$ and $\Gamma_{h\to \gamma\gamma}$.  We see the partial width of $h\to \phi \phi$ could be sizable in the parameter region where the diphoton width is enhanced significantly. For the benchmark (blue star) with $m_h = 126 \ \rm GeV$, $m_{\phi} = 92 \ \rm GeV$ and $c_{\phi} = 1$, a Higgs-to-diphoton enhancement of ~60\% leads to a partial decay width $\Gamma_{h\to \phi\phi} \sim 0.5\times  \Gamma^{\rm SM}_{h\to \tau \bar \tau}$. Recall BR$_{\rm SM}(h\to\tau \bar \tau)\approx 6.4$\% for a 126 GeV SM Higgs, so such an exotic decay mode potentially can lead to significant implications for colliders. A cautionary remark is that the exotic partial width becomes very small if both mediators in the decay are only allowed to be off-shell (i.e., if $m_\phi > m_h$), in which case our method may not be applicable. But, such a scenario usually requires a larger fine-tuning to escape theoretical and experimental constraints, given similar modification strength for the $h\gamma\gamma$ coupling.

\pp
{\bf LHC Study} -- We first consider the discovery potential at the LHC. Since each mediator decays into $\tau$ plus $\nu_\tau$, the signature is $2\tau+\slash p_{\rm T}$, which is very similar to the SM $h\to  \tau  \tau$ decay. In this work we only consider the vector-boson fusion (VBF) production of the Higgs, which has the best signal-to-background ratio in the SM $h\to \tau \bar \tau$ search \cite{Chatrchyan:2012vp}. In particular, we use the dileptonic channel $2\tau_l+\slash p_{\rm T}$ to illustrate the LHC sensitivity (with $l= e, \mu$). It is well-known that the VBF selection cuts have contaminations from the gluon fusion channel, which we include in our simulations.

In the simulation, the events are generated for the $\sqrt{s} = 14 \ \rm TeV$ LHC with MadGraph5 \cite{Alwall:2011uj} and are showered with Pythia6 \cite{Sjostrand:2006za}. Here the contributions from underlying events and pile-up are also turned on, with the pile-up cross section taken to be $0.25$ mb.  
Jet clustering is done in FastJet \cite{Cacciari:2011ma} using the anti-$k_t$ algorithm with a cone size of $R = 0.5$.  Lepton isolation requires the net $p_{\rm T}$ of particles in a cone of size $R = 0.3$ to be less than 10\% of the lepton's $p_{\rm T}$.  
The cross sections of the processes with no Higgs involved are obtained from Pythia, scaled by the appropriate $k$-factor \cite{Campbell:1999ah,Hamberg:1990np,Broggio:2011bd,Dittmaier:2011ti}, while the cross sections of the other ones are simply scaled from the SM predictions by comparing with the SM $h\to \tau \bar \tau$ width. The selection cuts are summarized in Table~\ref{VBFCutsTable}.

\begin{table}
\begin{center}
  \begin{tabular}{l|  p{7cm}l} \hline \hline 

 Cut 1  & Two jets with $p_{\rm T} > 20$ GeV each, $m_{jj} > 650$ GeV,  
                     $|\Delta \eta | > 3.5$, and $\eta_1 \eta_2 < 0$. Total jet $H_T > 80$ GeV,
                    and no additional jets with $p_{\rm T} > 30$ GeV between forward jets. 
                    \\ \hline 
   Cut 2  & Two opposite-sign leptons, harder with $10 \ \mathrm{GeV} < p_{\rm T} < 20 \ \mathrm{GeV}$,
   softer with $10 \ \mathrm{GeV} < p_{\rm T} < 15 \ \mathrm{GeV}$
      $|\eta| < 2.3$ for electrons; $|\eta| < 2.1$ for muons. \\ \hline 
  Cut 3       &  Invariant lepton mass $m_{ll} < 20 \ \rm GeV$, 
  		  $\slash p_{\rm T} > 40 \ \mathrm{GeV}$\ .\\     \hline \hline
  \end{tabular}
\end{center}
\caption{Cuts for the LHC analysis. }
\label{VBFCutsTable}
\end{table}

\begin{figure}[t]
\centering
\includegraphics[width=0.35\textwidth]{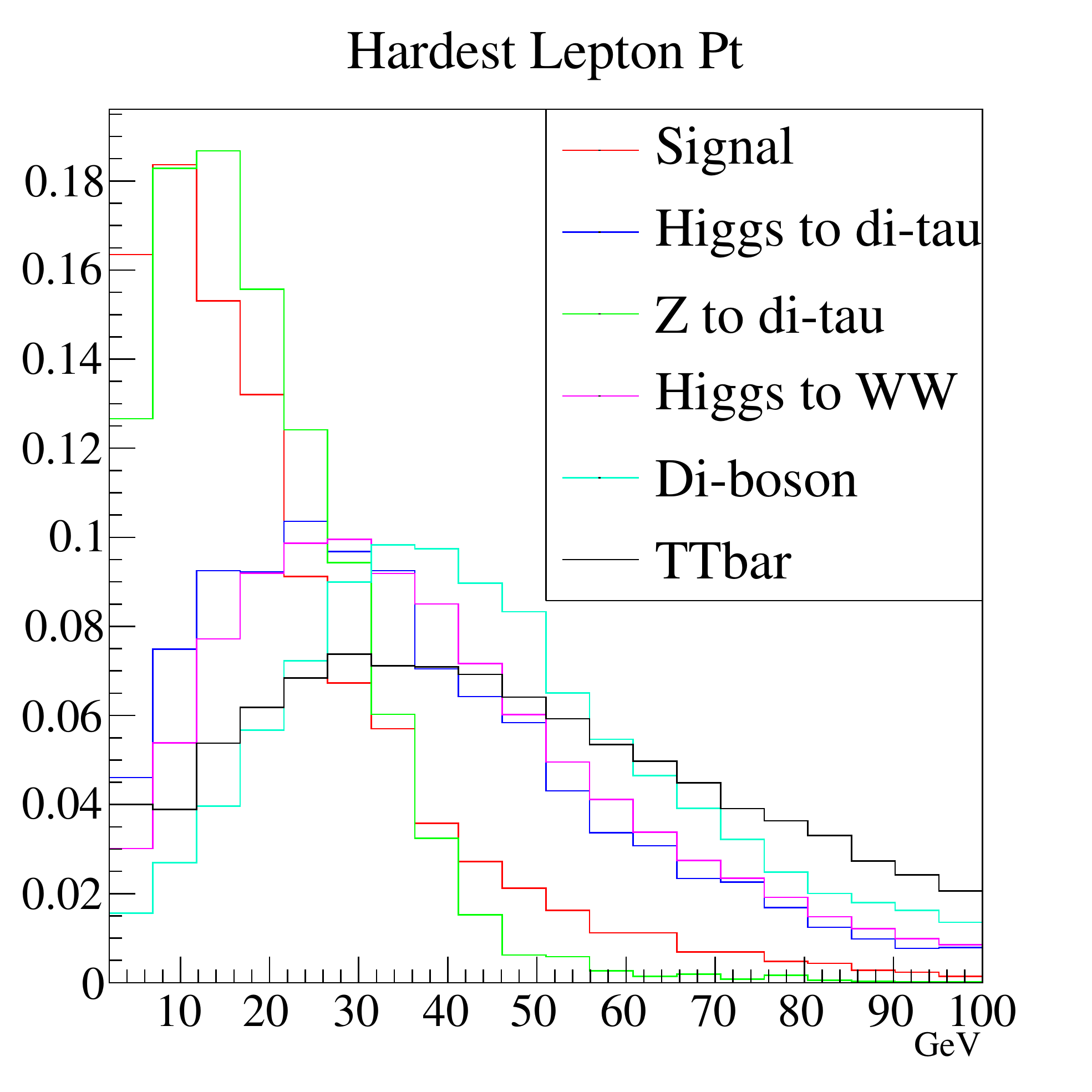}
\includegraphics[width=0.35\textwidth]{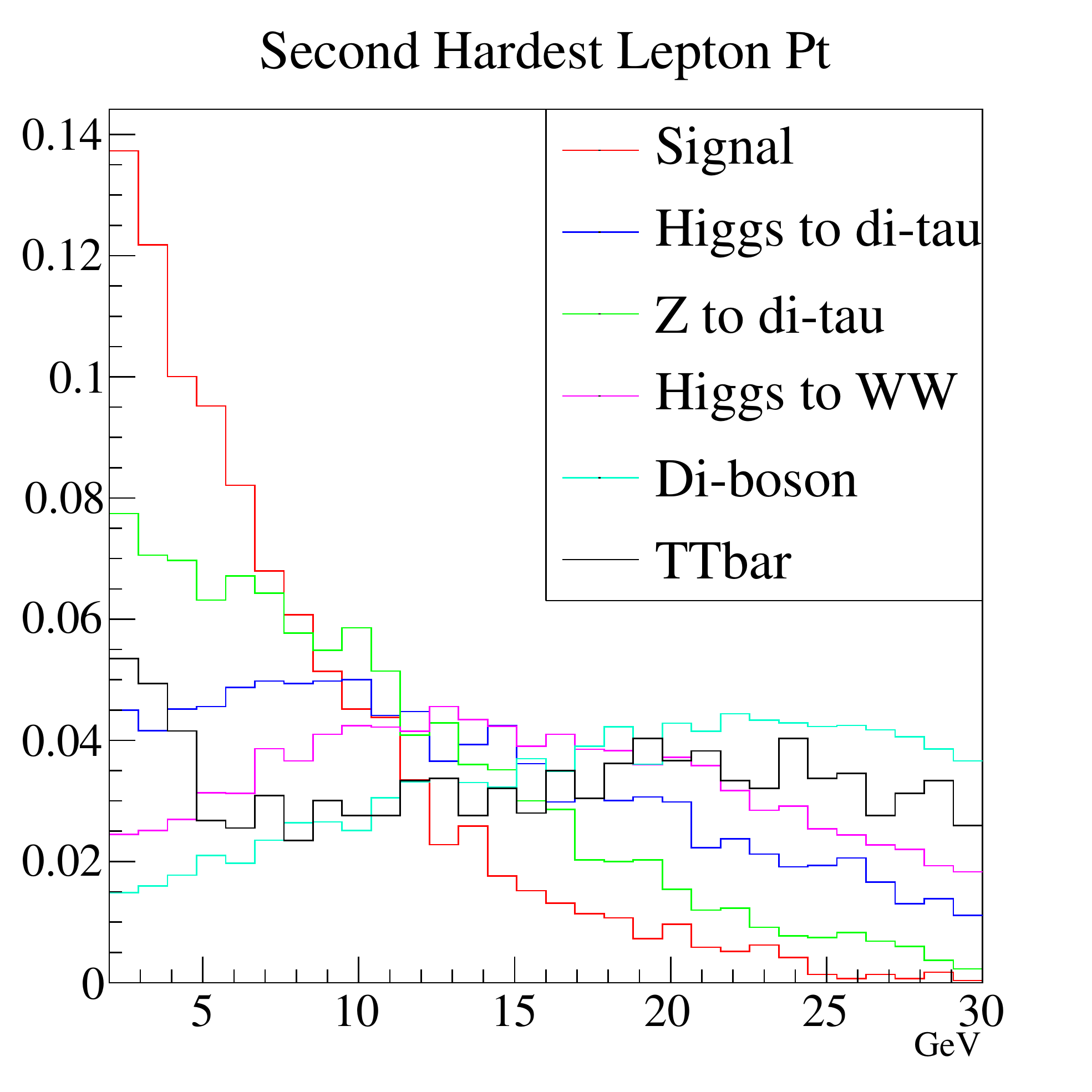}
\caption{Normalized distributions of lepton $p_{\rm T}$.}
\label{DiStauLeptonPtPlot}
\end{figure}

Fig.~\ref{DiStauLeptonPtPlot} shows the normalized $p_{\rm T}$ distributions of the two final-state leptons for the signal and the backgrounds.
We see that that the second hardest lepton for the signal has an extremely soft $p_{\rm T}$ distribution peaked at below 5 GeV, because the off-shell $\phi$ carries very little invariant mass. The feature exemplifies the challenge of making this discovery at the LHC, as the standard dilepton selection cut in Table~\ref{VBFCutsTable} eliminates most of the signal. One might consider forgoing the soft lepton and making a single lepton selection. However, in this case the signal is completely overwhelmed by the   $W + j$ background,  which is suppressed by dilepton selection and thus not included in our simulation. In the end, this analysis shares similar background with the SM $h\to  \tau_l  \tau_l$ search, which mainly includes $t\bar{t}$, di-boson and $(Z\to \tau \bar \tau)+2j$~\cite{Chatrchyan:2012vp}.  (On the other hand, the $(Z\to l^+l^-)+2j$ background is removed by the lower cut on $\slash p_{\rm T}$ and not included.)  In addition, the $h\to  \tau_l  \tau_l$ decay itself is a background for the exotic decay search. In addition to the VBF and dilepton selections, in Table~\ref{VBFCutsTable} we further require a maximum value in $m_{ll}$. For signal events, $m_{ll}$ tends to be small since one of the leptons is soft. An even stronger minimum cut on the missing energy does not help much after the VBF cut, since both the signal and background events left tend to have a relatively large missing energy.

\begin{table}[h]
\begin{center}
\begin{tabular}{|c|c|c|c|c|c|c|}
\hline
    14 TeV          & Signal & $h\to  \tau_l  \tau_l$ &  $h \to WW$ & $Z \to \tau_l \bar \tau_l$  & $t\bar{t}$ & Di-boson \\ \hline
    $\sigma$ (pb) & 0.06 & 0.11 & 0.27 &  0.72 & 8.0 & 0.17\\ \hline
Cut 1 & 1539 & 3041  & 6393 & 24757 & 9377 & 4421\\ \hline
Cut 2 & 33 & 66  & 74 &  327 & 11 & 13 \\ \hline 
Cut 3  & 16 & 2 & 16 & 40 & 2 & 4 \\ \hline
$S / \sqrt{B}$ & \multicolumn{6}{|c|}{ $\sim 2 \sigma$} \\ \hline

\end{tabular}

\end{center}
\caption{Cut flows in the LHC analysls.  Events produced are for $100 \ \mathrm{fb}^{-1}$ of data.  Gluon fusion contamination is included in the VBF selection.  Production cross sections are after preselection cuts which are different for different processes.}
\label{VBFCutFlow}
\end{table}

The cut flows for the signal and the backgrounds are summarized in Table \ref{VBFCutFlow}, where a luminosity of 100 fb$^{-1}$ is assumed.  We see that the search sensitivity is not too promising, unless new techniques for identifying very soft leptons are developed.

\pp
{\bf ILC Study} --
Next we demonstrate  that a Higgs factory such as the ILC could serve as a discovery machine for the Higgs decay to light EW mediators, 
even during its early run with $\sqrt{s} = 250$ GeV. To begin with, we assume a beam polarization $(P_{e^-},P_{e^+})=(0.8,-0.6)$ for the ILC and focus on the process ($l^{\pm}=e^\pm, \mu^\pm$)
\begin{eqnarray}
e^+e^- \to Z (h \to \phi\phi \to \tau \bar \tau \nu_\tau \bar \nu_\tau) \to l^+l^-l^+l^- + \slash p_T \ .
\end{eqnarray}
This decay topology provides an extremely clean laboratory, with the  main background being  $Z+ (Z/h\to \tau \bar \tau)$. Tri-boson production, $ZWW$, with all of them decaying leptonically, is subdominant and also included.

\begin{table}
\begin{center}
  \begin{tabular}{l| p{7cm} l} \hline \hline 
 Cut 1  & Three leptons $l_i$ ($i=1,2,3$),   with $|\cos \theta_{l_i}| < 0.99$, $E_{l_i}>3$~GeV and $E_{l_3} < 20$~GeV. Fourth-lepton (with $|\cos \theta_{l_4}| < 0.99$ and $E_{l_4} > 10$~GeV) veto.     \\ \hline 
   Cut 2  & $m_{l_1l_2} = 91.2 \pm 5$ GeV, $|\cos \theta_{l_1 l_2}| < 0.8$      \\ 
   \hline 
  Cut 3   & $\slash p_T > 70$~GeV \\
                   \hline 
Cut 4 & 125 GeV $< m_h^{\rm rec} <$ 150 GeV \\       \hline \hline
  \end{tabular}
\end{center}
\caption{Cuts for the ILC analysis. $\theta_{l_i}$ and $\theta_{l_1 l_2}$ are polar angles of the leptons and the reconstructed $Z$ boson ($l_1+l_2$) w.r.t. the beam, respectively. $m_h^{\rm rec}$ is the Higgs recoil mass.  \label{ILC_cuts}}
\end{table}

\begin{figure}[t]
\centering
\includegraphics[width=0.35\textwidth]{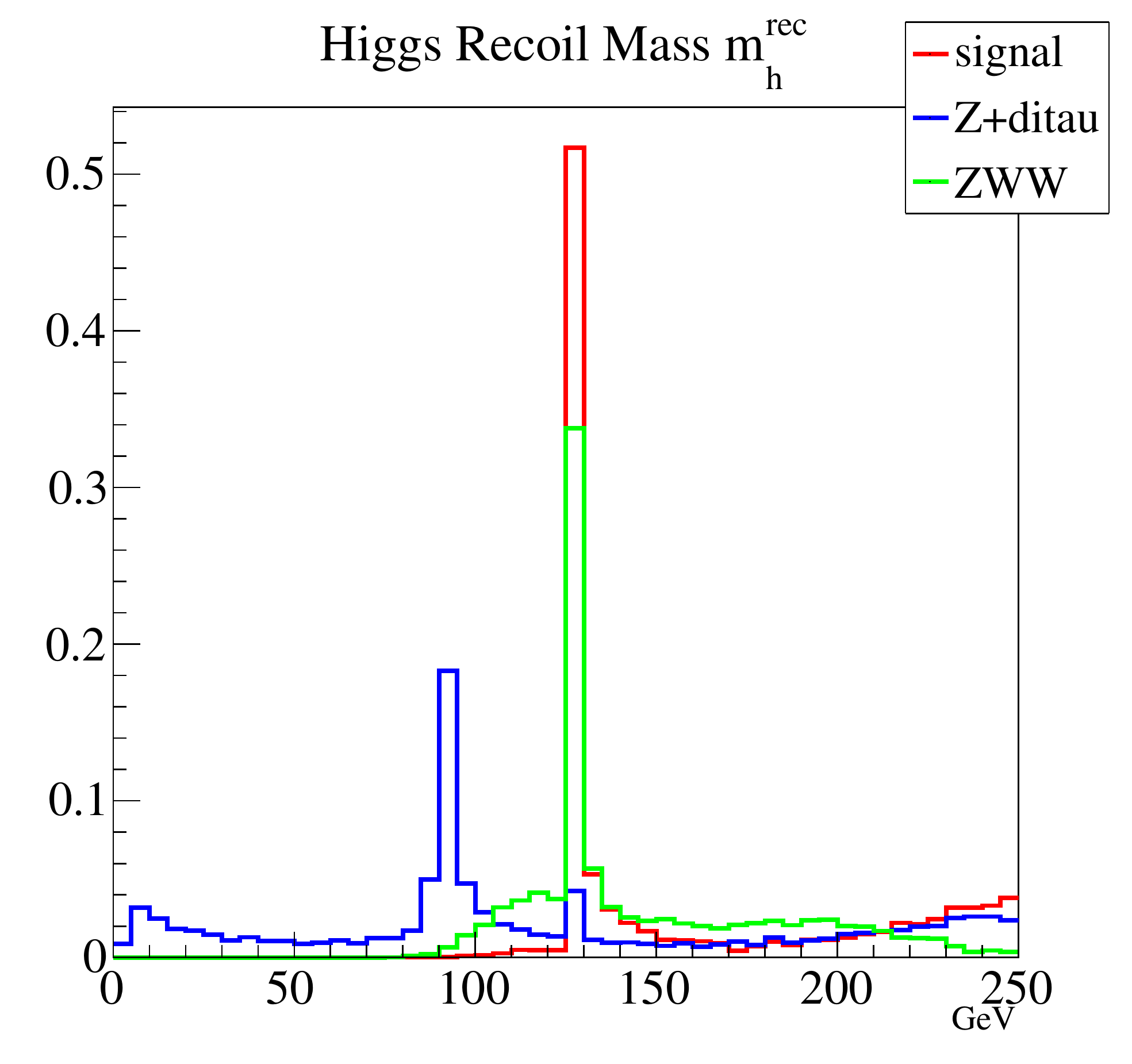}
\caption{Normalized distributions of the Higgs recoil mass. }
\label{ILC_Invm_MET}
\end{figure}

The selection cuts are summarized in Table~\ref{ILC_cuts}. Given that the fourth charged lepton, the one from the off-shell mediator decay, is extremely soft, we use a three-lepton selection and introduce a hard fourth-lepton veto to enhance the signal.  In addition, due to a relatively small $\sqrt{s}$, the angular distribution of the $Z$ boson in $e^+e^- \to Zh$ is flat in $\cos\theta$ \cite{Djouadi:2005gi}, while the $Z$ bosons in the $Z\tau \bar \tau$ events are more forward because most of the $Z\tau \bar \tau$ events are from the di-$Z$ production, which proceeds via $t$-channel processes. So in Table~\ref{ILC_cuts} we require $|\cos \theta_{l_1l_2}| < 0.8$ to suppress the $Z\tau \bar \tau$ background. Further suppression is achieved by demanding $\slash{p}_T > 50$ GeV.

In Fig.~\ref{ILC_Invm_MET} we show  the normalized distribution of the Higgs recoil mass, $m_h^{\rm rec} \equiv \sqrt{s-2 \sqrt{s} E_{l_1l_2} +m_{l_1l_2}^2}$, for both signal and backgrounds, where the peak at the $m_h=126$ GeV for the signal is difficult to miss. This figure demonstrates the advantage of knowing the center-of-mass energy in a lepton collider such as the ILC: the Higgs mass can be reconstructed precisely even with missing particles in the final state. Our last cut in Table~\ref{ILC_cuts} utilizes $m_h^{\rm rec}$ to cut away the diboson background $Z+(Z\to \tau \bar \tau)$, which is peaked at $m_Z$ in Fig.~\ref{ILC_Invm_MET}. It is also interesting to see that both $Z\tau \bar \tau$ and $ZWW$ backgrounds receive contributions from $Z+(h\to WW/\tau \bar \tau)$ processes.

\begin{table}[h]
\begin{center}
\begin{tabular}{|c|c|c|c|}
\hline
     $\sqrt{s}=250$ GeV  & signal & $Z \tau \bar \tau$  &  $ZWW$   \\ \hline    
         Xsection (fb) & $0.93$ & 27.81 &  0.02  \\ \hline 
Events  & 10000 & 10000& 10000  \\ \hline
Cut 1   &2420  & 1854 & 1404  \\ \hline
Cut 2   & 1272 & 575 & 329  \\ \hline
Cut 3   &821&93& 258  \\ \hline
Cut 4   & 820 &  3& 255  \\ \hline
$S / \sqrt{B}$   &  \multicolumn{3}{|c|}{$\sim 5.2 \sigma$}  \\ \hline
\end{tabular}
\end{center}
\caption{Cut flows in the ILC analysis. $S / \sqrt{B}$ is calculated for 40 fb$^{-1}$ of data.}\label{cutflowilc}
\end{table}

The cut flows for the signal and the backgrounds are summarized in Table~\ref{cutflowilc}. 
For the benchmark that we are considering, the signal cross section is about half of the SM $h\to \tau \bar \tau$. We see that a $S/ \sqrt {B}\approx 5\sigma$ discovery can be made with about 40 fb$^{-1}$ of data at $\sqrt{s}=250$ GeV.

\pp
{\bf Conclusion} -- We have shown that light EW mediators contributing to a strongly enhanced Higgs-to-diphoton width could show up in exotic Higgs decays. We then proposed using such decays to uncover the mediators and explore their couplings to the Higgs. In general one of the mediators in the exotic Higgs decay is far off-shell and its decay products are very soft, which makes it difficult to search for at the LHC, unless new techniques for identifying soft tracks are introduced. On the other hand, such discoveries can be made with a relatively small amount of data at the ILC with $\sqrt{s}=250$ GeV. Obviously a detailed comparison between the discovery reaches at the LHC and the ILC, as well as generalizations to other types of mediators and decay final states, are warranted, and will be reported elsewhere~\cite{GLLM}.

\begin{center}
{\bf Acknowledgments}
\end{center}

We acknowledge discussions with Claudio Campagnari, Tom Danielson, Joe Lykken, Vyacheslav Krutelyov, Jim Olsen, Yanjun Tu and El Carlos Wagner. S.G. and T.L. are supported in part by DOE under grant DE-FG02-91ER40618, and S.G. is supported in part by a Simons Foundation Fellowship, 229624, to Steven Giddings. I.L. is supported in part by DOE under Contract No. DE-AC02-06CH11357 (ANL) and No. DE-FG02-91ER40684 (Northwestern), and by the Simons Foundation under award No. 230683. Work at KITP is supported by the National Science Foundation under Grant No. PHY11-25915.

\end{document}